\documentclass[utf8]{frontiersSCNS} 

\usepackage{url,hyperref,microtype,subcaption}
\usepackage[onehalfspacing]{setspace}

\def\keyFont{\fontsize{8}{11}\helveticabold }
\def\firstAuthorLast{Mapelli} 
\def\Authors{Michela Mapelli\,$^{1,2,3}$}
\def\Address{$^{1}$Physics and Astronomy Department Galileo Galilei, University of Padova, Vicolo dell'Osservatorio 3, I--35122, Padova, Italy \\
  $^{2}$INFN--Padova, Via Marzolo 8, I--35131 Padova, Italy\\
$^{3}$INAF--Osservatorio Astronomico di Padova, Vicolo dell'Osservatorio 5, I--35122, Padova, Italy}
\def\corrAuthor{Michela Mapelli}

\def\corrEmail{michela.mapelli@unipd.it}

\begin{document}
\onecolumn
\firstpage{1}

\title[BBH mergers: formation and populations]{Binary black hole mergers: formation and populations} 

\author[\firstAuthorLast]{\Authors} 
\address{} 
\correspondance{} 

\extraAuth{}

\maketitle

\begin{abstract}
We review the main physical processes that lead to the formation of stellar binary black holes (BBHs) and to their merger. BBHs can form from the isolated evolution of massive binary stars. The physics of core-collapse supernovae and the process of common envelope are two of the main sources of uncertainty about this formation channel. Alternatively, two black holes can form a binary by dynamical encounters in a dense star cluster. The dynamical formation channel leaves several imprints on the mass, spin and orbital properties of BBHs.
\tiny
 \keyFont{ \section{Keywords:} stars: black holes -- black hole physics -- Galaxy: open clusters and associations: general -- stars: kinematics and dynamics -- gravitational waves} 
\end{abstract}

\section{Black hole formation from single stars: where we stand now}
About four years ago, the LIGO detectors obtained the first direct detection of gravitational waves, GW150914 \citep{abbottGW150914,abbottO1,abbottastrophysics}, associated with the merger of two black holes (BHs). This event marks the dawn of gravitational wave astronomy: we now know that binary black holes (BBHs) exist, can reach coalescence by gravitational wave emission, and are composed of BHs with mass ranging from a few to a few ten solar masses\footnote{The detection of GW190521 \citep{abbottGW190521,abbottGW190521astro} was reported after this review was sent to the press.}. Here, we review the main physical processes that lead to the formation of BBHs and to their merger.  We restrict our attention to stellar-born BHs. As to primordial BHs, which might form from gravitational instabilities in the early Universe, we refer the reader to \cite{carr2016,belotsky2019}, and references therein. Before we start discussing binaries, we must briefly summarize the state-of-the-art knowledge about stellar-origin BHs: this is a necessary step to understand their pairing mechanisms.

Stellar-mass BHs are thought to be the final outcome of the evolution of a massive star (zero-age main sequence mass $m_{\rm ZAMS}\gtrsim{}20$ M$_\odot$). Hence the mass of the BH should be affected by the two main processes that influence the evolution of a single star: i) mass loss by stellar winds and ii) the final collapse.

\subsection{Stellar winds}
Hot ($>10^4$ K) massive stars ($m_{\rm ZAMS}\gtrsim{}30$ M$_\odot$) lose a non-negligible fraction of their mass by line-driven winds. This process depends on metallicity ($Z$): the mass-loss rate by stellar winds can be described as $\dot{m}\propto{}Z^\beta$, where $Z$ is the absolute metallicity (see e.g. \citealt{vink2001} and references therein). The most recent models suggest that $\beta{}$ is not constant, but depends at least on the luminosity of the star \citep{graefener2008,vink2011,chen2015}: the closer the luminosity $L_\ast$ is to the Eddington value $L_{\rm Edd}$, the higher the mass loss, basically cancelling the dependence on metallicity when $L_\ast\gtrsim{}L_{\rm Edd}$.

In single stars, stellar winds uniquely determine the final mass of the star at the onset of collapse. If we consider a star with $m_{\rm ZAMS}=90$ M$_\odot$ and metallicity $Z=0.02$ (i.e. approximately solar), its final mass will be only $m_{\rm fin}\sim{}30$ M$_\odot$; while the same star with $Z<0.0002$ has $m_{\rm fin}\gtrsim{}0.8\,{}m_{\rm ZAMS}$. The final mass of a star $m_{\rm fin}$ is the strongest upper limit to the mass of the BH. 

\subsection{Core-collapse supernovae}
 Several models in the literature try to predict the final outcome of a core-collapse supernova (SN). 
 For example, \cite{fryer1999} and \cite{fryer2001} suggest that if the final mass of the star is $m_{\rm fin}\gtrsim{}40$ M$_\odot$, the fate of the star is to collapse to a BH directly, without supernova, because the binding energy of the outer stellar layers is too big to be overcome by the explosion. \cite{fryer2012} elaborate on these early results proposing that the mass of the compact object depends not only on $m_{\rm fin}$ but also on the final mass of the carbon-oxygen core. Alternatively, \cite{oconnor2011} proposed the role of the compactness parameter $\xi{}_M=\frac{M/{\rm M}_\odot}{R(\leq{}M)/1000\,{}{\rm km}}$: if the compactness is small (e.g. $\xi{}_{2.5}\le{}0.2-0.4$), the SN explosion is successful, otherwise we expect the star to collapse directly. All of these simplified models as well as more sophisticated ones (e.g. \citealt{ertl2016}) point toward a similar direction: if the star ends its life with a large final mass, its carbon-oxygen core grows larger, its compactness is generally higher, and so on. Hence, we expect that metal-poor stars, which retain a larger fraction of their mass to the very end and develop larger cores, are more likely to collapse to BHs directly, producing larger BHs (e.g. \citealt{mapelli2009,mapelli2010,zampieri2009,belczynski2010}). This simplified picture seems to agree with observations, but must be taken with several grains of salt: we need a vigorous step forward in core-collapse SN simulations and theoretical models, before we can draw robust conclusions (e.g. \citealt{burrows2018}).


 \subsection{Pair instability}
 Core-collapse SNe are not the only mechanism that can end the life of a massive star. When the helium core of a star grows to $\ge{}60$ M$_\odot$ and the central temperature reaches $\sim{}10^9$~K, electron and positron pairs are produced at an efficient rate, leading to a softening of the equation of state. The star undergoes pair instability (PI, \citealt{ober1983, bond1984, heger2003, woosley2007}): oxygen, neon and silicon are burned explosively and the entire star is disrupted leaving no remnant, unless its helium core is $\ge{}130$~M$_\odot$. In the latter case, the gravity of the outer layers is so big that the star collapses to a massive BH directly as an effect of PI  \citep{heger2003}. Smaller helium cores ($\sim{}30-60$ M$_\odot$) are associated with a less dramatic manifestation of PI: the softened equation of state drives oscillations of the core (pulsational PI, \citealt{barkat1967, woosley2007,chen2014,yoshida2016}); during each oscillation the star sheds some mass till it finds a new equilibrium to a lower core mass, but leaves a BH smaller than expected without pulsational PI \citep{woosley2017,woosley2019,belczynski2016pair,spera2017,marchant2019,stevenson2019,renzo2020}.

\begin{figure}
\begin{center}
\includegraphics[width=15cm]{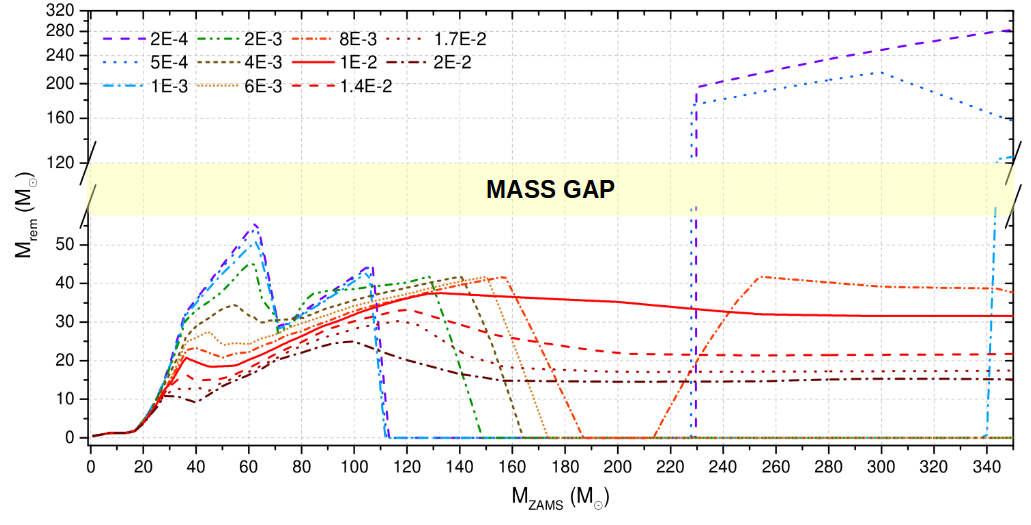}
\end{center}
\caption{Predicted compact object mass ($M_{\rm rem}$) as a function of the zero-age main-sequence (ZAMS) mass of the progenitor star ($M_{\rm ZAMS}$) for 11 different metallicities, ranging from $Z=2\times{}10^{-4}$ to $Z=2\times{}10^{-2}$, as shown in the legend. The yellow area highlights the pair-instability mass gap. These models are obtained with the SEVN population synthesis code \citep{spera2019}, using PARSEC evolutionary tracks \citep{bressan2012} and the delayed model from \cite{fryer2012}. See  \cite{spera2017} for details.}\label{fig:PI}
\end{figure}

 From the combination of PI, core-collapse SNe and stellar-wind mass loss prescriptions, 
 we expect the mass spectrum of BHs to behave roughly as shown in Figure~\ref{fig:PI}. In particular, PI is expected to carve a gap in the mass spectrum of BHs between $\sim{}50(-10,+20)$ M$_\odot$ and $\approx{}120-130$~M$_\odot$. The uncertainty on this mass gap is mainly connected with uncertainties nuclear reaction rates \citep{farmer2019}, on the collapse of the residual hydrogen envelope  and on the role of stellar rotation \citep{mapelli2020}. Within this framework, we predict a reasonable mass range for stellar-origin BHs to be $\sim{}3-65$ M$_\odot$ (assuming the most conservative value for the lower edge of pair-instability mass gap). If exotic metal-poor stars exist with mass $m_{\rm ZAMS}>250$ M$_\odot$, these might directly collapse to intermediate-mass BHs (IMBHs) with mass $>100$ M$_\odot$.

\section{Binary BH formation in isolation}
The scenario highlighted in the previous section assumes that the progenitor star is single. But gravitational waves have shown the existence of BBHs with a very short orbital separation: the initial separation of a BBH must be of the order of few ten solar radii for the BBH to merge within a Hubble time by gravitational-wave emission. This challenges our understanding of binary star evolution. A close binary star undergoes several physical processes during its life, which can completely change its final fate (see e.g. \citealt{eggleton2006}). The most important processes include mass transfer and common envelope, tides and natal kicks \citep{hurley2002}.

Mass transfer and common envelope are crucial in this regard. After main sequence, a massive star can develop a stellar radius as large as several thousand solar radii. Hence, if this star is member of a binary system and its orbital separation is of few hundred to few thousand solar radii, the binary system undergoes Roche lobe overflow and possibly common envelope \citep{ivanova2013}. If common envelope occurs between a BH and a giant companion star, the BH and the core of the giant star orbit about each other surrounded by the giant's envelope: they feel a strong gas drag from the envelope and lose kinetic energy, inspiralling about each other. This transfers thermal energy to the envelope, which might trigger the ejection of the envelope. If the envelope is not ejected, the binary system merges prematurely giving birth to a single BH. In contrast, if the envelope is ejected, the final binary system is composed of the BH and the core of the giant. Because of the spiral-in, the final semi-major axis of the binary is just few solar radii, much smaller than the initial one. If the naked core collapses to a BH without receiving a strong natal kick, the system becomes a BBH with a short orbital period, able to merge within a Hubble time. Unfortunately, our understanding of common envelope is still poor (see \citealt{fragos2019} for a recent simulation) and this uncertainty heavily affects our knowledge of BBH demography. The left-hand panel of Figure~\ref{fig:cartoon} is a schematic view of the isolated binary evolution channel through common envelope.

Several alternative scenarios to common envelope have been proposed \citep{marchant2016,demink2016,mandel2016}. For example, in the over-contact binary evolution, \cite{marchant2016} show that, when two massive stars in a tight binary are fast rotators,   they remain fully mixed as a result of their tidally induced high spin; in this case, the binary avoids premature merger even if it is overfilling its Roche lobe and might evolve into a tight BBH.





The isolated binary evolution scenario has several characteristic signatures. In the {\it common envelope} isolated binary evolution scenario, the {\it masses} of the two BHs span from $\sim{}3$ M$_\odot$ up to $\sim{}45$ M$_\odot$ (see e.g. \citealt{giacobbo2018b}) and the {\it mass ratios} are preferentially close to 1 (although all mass ratios $q=m_2/m_1\gtrsim{}0.1$ are possible, see e.g. \citealt{giacobbo2018b}). Most processes in binary evolution tend to produce aligned {\it spins} (e.g. \citealt{rodriguez2016b}), while the magnitude of the spin is basically unconstrained (but see \citealt{qin2018,qin2019,fuller2019}  for some recent attempts to quantify spins). Mass transfer episodes and gravitational-wave decay are expected to efficiently damp {\it eccentricity}, so that almost all isolated binaries have near zero eccentricity in the LIGO-Virgo band. Finally, local merger rate densities span from a few to few thousand events Gpc$^{-3}$ yr$^{-1}$, depending on the details of common envelope and natal kicks (e.g. \citealt{dominik2013,belczynski2016,mapelli2017,mapelli2018,giacobbo2018b,giacobbo2020,neijssel2019,tang2020,santoliquido2020}). 
The scenarios which include alternatives to common envelope predict an even stronger prevalence of systems with $q\sim{}1$, a preferred mass range $\sim{}25-60$ M$_\odot$ \citep{marchant2016}, high and aligned spins, zero eccentricity in the LIGO-Virgo band,  and long delay times ($\gtrsim{}3$ Gyr, \citealt{demink2016}). Local merger rate densities are expected to be $\sim{}10$ Gpc$^{-3}$ yr$^{-1}$ \citep{mandel2016}, with large uncertainties.






\section{Binary BH formation in star clusters}
Star clusters are among the densest places in the Universe. There is a plethora of star clusters, with their distinguishing features: i) globular clusters \citep{gratton2019} are old ($\sim{}12$ Gyr) and massive systems ($\sim{}10^{4-6}$ M$_\odot$), ii) nuclear star clusters can be even more massive ($\sim{}10^7$ M$_\odot$) and lie at the centre of many galaxies, in some cases coexisting with the supermassive BH \citep{neumayer2020}, iii) open clusters and young star clusters \citep{portegieszwart2010} are generally less massive (up to $\sim{}10^5$ M$_\odot$) and short lived (less than a few Gyr), but are the main birthplace of massive stars in the local Universe \citep{lada2003}.

The central density of star clusters is sufficiently high ($\gtrsim{}10^3$ stars pc$^{-3}$) and  their typical velocity dispersion sufficiently low (from a few to a few tens of km s$^{-1}$, possibly with the exception of nuclear star clusters) that their central two-body relaxation time \citep{spitzer1987} is shorter than their lifetime. This has one fascinating implication: the orbits of stars and binary stars in a star cluster are constantly perturbed by dynamical encounters with other cluster members. This process  affects the formation and the evolution of binary BHs  in multiple ways (e.g. \citealt{portegieszwart2000}).

{\emph{Dynamical exchanges}} occur when a binary system interacts with a single stellar object and the latter replaces one of the members of the binary. We have known for a long time that massive objects are more likely to acquire companions by dynamical exchanges \citep{hills1980}. Since BHs are among the most massive objects in a star cluster, they are very efficient in forming new binaries through exchanges (e.g. \citealt{ziosi2014}).

During a three-body encounter, a binary star exchanges a fraction of its internal energy with the third body. If the binary is particularly tight (hard binary), such encounters tend to harden the binary star, i.e. to increase its binding energy by reducing its semi-major axis ({\emph{dynamical hardening}}). In the case of a BBH, this hardening might speed up the merger, because it drives the semi-major axis of the BBH in the regime where orbital decay by gravitational waves becomes efficient (see e.g. Figure~10 of \citealt{mapelli2018c}). On the other hand, the least massive BBHs can even be {\emph{ionized}}, i.e. split by strong dynamical encounters with massive intruders.

{\emph{Mergers of massive stars}} are common in dense young star clusters, because of the short dynamical friction timescale \citep{portegieszwart2010}. Under some assumptions, these mergers can lead to the formation of massive BHs ($m_{\rm BH}>60$ M$_\odot$), with mass in the pair-instability gap \citep{dicarlo2019a}. In star clusters, such massive BHs  can acquire a companion by dynamical exchanges, leading to the formation of BBHs in the mass gap. A fast sequence of stellar mergers in the dense core of a young star cluster (also known as {\emph{runaway collision}}, \citealt{portegieszwart2004,giersz2015}) might even lead to the formation of intermediate-mass BHs (IMBHs), i.e. BHs with mass $m_{\rm BH}>100$ M$_\odot$, especially at low metallicity \citep{mapelli2016}.

\begin{figure}
\begin{center}
\includegraphics[width=15cm]{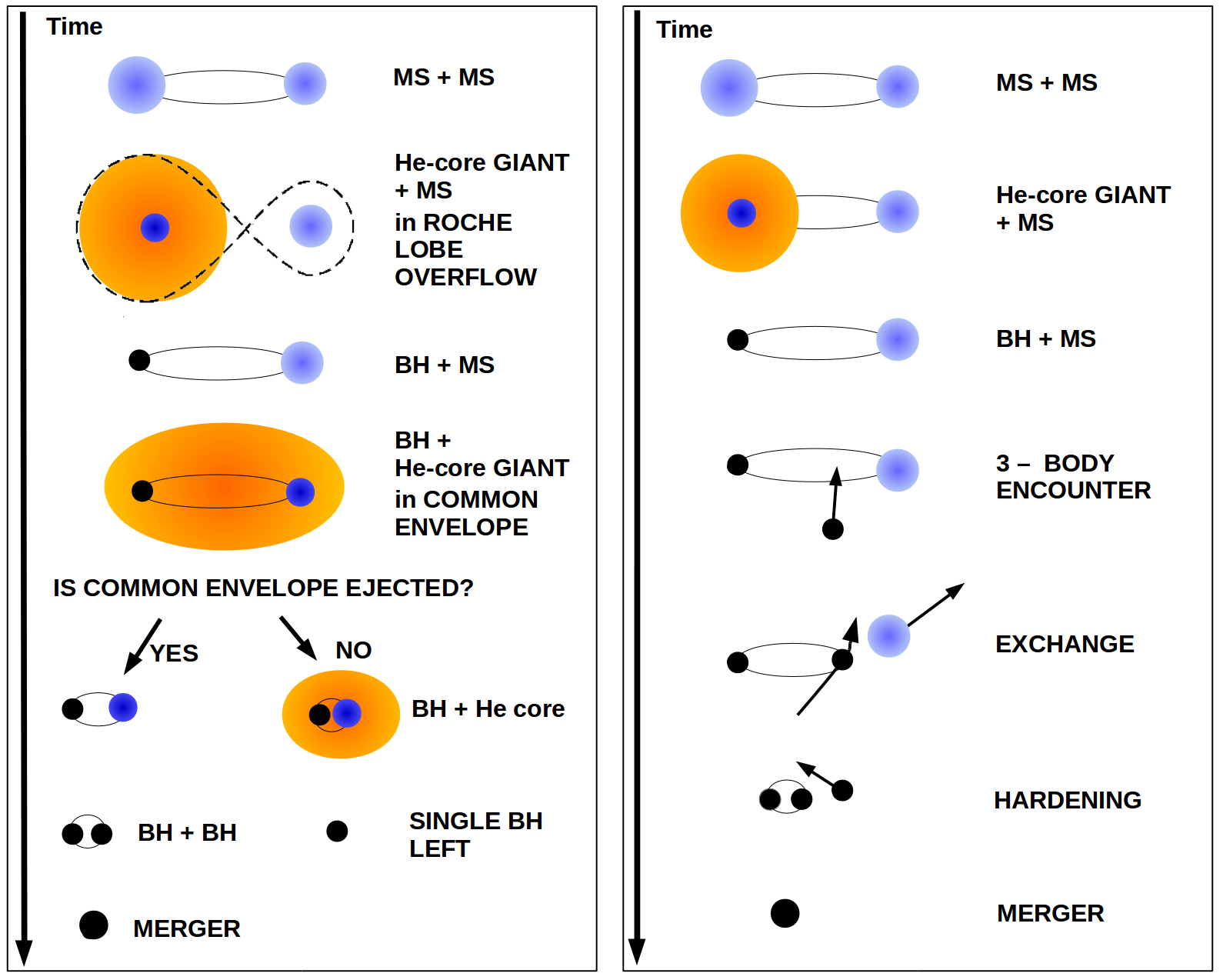}
\end{center}
\caption{Left-hand panel: cartoon of isolated BBH formation through common envelope; right-hand panel: cartoon of dynamical BBH formation in star clusters.}\label{fig:cartoon}
\end{figure} 

The dynamical processes we briefly summarized above (and in the right-hand panel of Figure~\ref{fig:cartoon}) leave a clear imprint on BBHs. First, dynamically formed BBHs extend to higher masses than isolated BBHs: they might even be in the pair-instability mass gap or in the IMBH regime \citep{dicarlo2019b,rodriguez2019}. Secondly, dynamical exchanges randomize the spin direction, leading to an isotropic distribution of BH spins. In contrast, isolated BBHs have a preference for aligned spins \citep{rodriguez2016b,gerosa2018}. Third, dynamics can trigger the merger of BBHs with non-zero eccentricity even in the LIGO-Virgo band \citep{samsing2014,samsing2018,rodriguez2018,zevin2019}. These signatures provide an unique opportunity to differentiate among the isolated and the dynamical formation channel when the number of gravitational-wave detections will be of the order of a few hundreds (e.g. \citealt{zevin2017,bouffanais2019}).














\section*{Author Contributions}

MM collected the material and wrote the review.

\section*{Funding}

MM acknowledges financial support from the European Research Council for the ERC Consolidator grant DEMOBLACK, under contract no. 770017.

\bibliographystyle{frontiersinSCNS_ENG_HUMS}

\bibliography{biblio}

\begin{thebibliography}{81}
\providecommand{\natexlab}[1]{#1}
\expandafter\ifx\csname urlstyle\endcsname\relax
  \providecommand{\doi}[1]{doi:\discretionary{}{}{}#1}\else
  \providecommand{\doi}{doi:\discretionary{}{}{}\begingroup
  \urlstyle{rm}\Url}\fi
\providecommand{\selectlanguage}[1]{\relax}
\providecommand{\bibAnnoteFile}[1]{%
  \IfFileExists{#1}{\begin{quotation}\noindent\textsc{Key:} #1\\
  \textsc{Annotation:}\ \input{#1}\end{quotation}}{}}
\providecommand{\bibAnnote}[2]{%
  \begin{quotation}\noindent\textsc{Key:} #1\\
  \textsc{Annotation:}\ #2\end{quotation}}

\bibitem[{{Abbott} et~al.(2016{\natexlab{a}}){Abbott}, {Abbott}, {Abbott},
  {Abernathy}, {Acernese}, {Ackley} et~al.}]{abbottastrophysics}
{Abbott}, B.~P., {Abbott}, R., {Abbott}, T.~D., {Abernathy}, M.~R., {Acernese},
  F., {Ackley}, K., et~al. (2016{\natexlab{a}}).
\newblock {Astrophysical Implications of the Binary Black-hole Merger
  GW150914}.
\newblock \emph{\apjl} 818, L22.
\newblock \doi{10.3847/2041-8205/818/2/L22}
\bibAnnoteFile{abbottastrophysics}

\bibitem[{{Abbott} et~al.(2016{\natexlab{b}}){Abbott}, {Abbott}, {Abbott},
  {Abernathy}, {Acernese}, {Ackley} et~al.}]{abbottO1}
{Abbott}, B.~P., {Abbott}, R., {Abbott}, T.~D., {Abernathy}, M.~R., {Acernese},
  F., {Ackley}, K., et~al. (2016{\natexlab{b}}).
\newblock {Binary Black Hole Mergers in the First Advanced LIGO Observing Run}.
\newblock \emph{Physical Review X} 6, 041015.
\newblock \doi{10.1103/PhysRevX.6.041015}
\bibAnnoteFile{abbottO1}

\bibitem[{Abbott et~al.(2016)Abbott, Abbott, Abbott, Abernathy, Acernese,
  Ackley et~al.}]{abbottGW150914}
Abbott, B.~P., Abbott, R., Abbott, T.~D., Abernathy, M.~R., Acernese, F.,
  Ackley, K., et~al. (2016).
\newblock Observation of gravitational waves from a binary black hole merger.
\newblock \emph{Phys. Rev. Lett.} 116, 061102.
\newblock \doi{10.1103/PhysRevLett.116.061102}
\bibAnnoteFile{abbottGW150914}

\bibitem[{{Abbott} et~al.(2020{\natexlab{a}}){Abbott}, {Abbott}, {Abraham},
  {Acernese}, {Ackley}, {Adams} et~al.}]{abbottGW190521}
{Abbott}, R., {Abbott}, T.~D., {Abraham}, S., {Acernese}, F., {Ackley}, K.,
  {Adams}, C., et~al. (2020{\natexlab{a}}).
\newblock {GW190521: A Binary Black Hole Merger with a Total Mass of 150
  M$_{\odot}$}.
\newblock \emph{\prl} 125, 101102.
\newblock \doi{10.1103/PhysRevLett.125.101102}
\bibAnnoteFile{abbottGW190521}

\bibitem[{{Abbott} et~al.(2020{\natexlab{b}}){Abbott}, {Abbott}, {Abraham},
  {Acernese}, {Ackley}, {Adams} et~al.}]{abbottGW190521astro}
{Abbott}, R., {Abbott}, T.~D., {Abraham}, S., {Acernese}, F., {Ackley}, K.,
  {Adams}, C., et~al. (2020{\natexlab{b}}).
\newblock {Properties and Astrophysical Implications of the 150
  M$_{{\ensuremath{\odot}}}$ Binary Black Hole Merger GW190521}.
\newblock \emph{\apjl} 900, L13.
\newblock \doi{10.3847/2041-8213/aba493}
\bibAnnoteFile{abbottGW190521astro}

\bibitem[{{Barkat} et~al.(1967){Barkat}, {Rakavy}, and {Sack}}]{barkat1967}
{Barkat}, Z., {Rakavy}, G., and {Sack}, N. (1967).
\newblock {Dynamics of Supernova Explosion Resulting from Pair Formation}.
\newblock \emph{\prl} 18, 379--381.
\newblock \doi{10.1103/PhysRevLett.18.379}
\bibAnnoteFile{barkat1967}

\bibitem[{{Belczynski} et~al.(2010){Belczynski}, {Bulik}, {Fryer}, {Ruiter},
  {Valsecchi}, {Vink} et~al.}]{belczynski2010}
{Belczynski}, K., {Bulik}, T., {Fryer}, C.~L., {Ruiter}, A., {Valsecchi}, F.,
  {Vink}, J.~S., et~al. (2010).
\newblock {On the Maximum Mass of Stellar Black Holes}.
\newblock \emph{\apj} 714, 1217--1226.
\newblock \doi{10.1088/0004-637X/714/2/1217}
\bibAnnoteFile{belczynski2010}

\bibitem[{{Belczynski} et~al.(2016{\natexlab{a}}){Belczynski}, {Heger},
  {Gladysz}, {Ruiter}, {Woosley}, {Wiktorowicz} et~al.}]{belczynski2016pair}
{Belczynski}, K., {Heger}, A., {Gladysz}, W., {Ruiter}, A.~J., {Woosley}, S.,
  {Wiktorowicz}, G., et~al. (2016{\natexlab{a}}).
\newblock {The effect of pair-instability mass loss on black-hole mergers}.
\newblock \emph{\aap} 594, A97.
\newblock \doi{10.1051/0004-6361/201628980}
\bibAnnoteFile{belczynski2016pair}

\bibitem[{{Belczynski} et~al.(2016{\natexlab{b}}){Belczynski}, {Holz}, {Bulik},
  and {O'Shaughnessy}}]{belczynski2016}
{Belczynski}, K., {Holz}, D.~E., {Bulik}, T., and {O'Shaughnessy}, R.
  (2016{\natexlab{b}}).
\newblock {The first gravitational-wave source from the isolated evolution of
  two stars in the 40-100 solar mass range}.
\newblock \emph{\nat} 534, 512--515.
\newblock \doi{10.1038/nature18322}
\bibAnnoteFile{belczynski2016}

\bibitem[{{Belotsky} et~al.(2019){Belotsky}, {Dokuchaev}, {Eroshenko},
  {Esipova}, {Khlopov}, {Khromykh} et~al.}]{belotsky2019}
{Belotsky}, K.~M., {Dokuchaev}, V.~I., {Eroshenko}, Y.~N., {Esipova}, E.~A.,
  {Khlopov}, M.~Y., {Khromykh}, L.~A., et~al. (2019).
\newblock {Clusters of Primordial Black Holes}.
\newblock \emph{European Physical Journal C} 79, 246.
\newblock \doi{10.1140/epjc/s10052-019-6741-4}
\bibAnnoteFile{belotsky2019}

\bibitem[{{Bond} et~al.(1984){Bond}, {Arnett}, and {Carr}}]{bond1984}
{Bond}, J.~R., {Arnett}, W.~D., and {Carr}, B.~J. (1984).
\newblock {The evolution and fate of Very Massive Objects}.
\newblock \emph{\apj} 280, 825--847.
\newblock \doi{10.1086/162057}
\bibAnnoteFile{bond1984}

\bibitem[{{Bouffanais} et~al.(2019){Bouffanais}, {Mapelli}, {Gerosa}, {Di
  Carlo}, {Giacobbo}, {Berti} et~al.}]{bouffanais2019}
{Bouffanais}, Y., {Mapelli}, M., {Gerosa}, D., {Di Carlo}, U.~N., {Giacobbo},
  N., {Berti}, E., et~al. (2019).
\newblock {Constraining the Fraction of Binary Black Holes Formed in Isolation
  and Young Star Clusters with Gravitational-wave Data}.
\newblock \emph{\apj} 886, 25.
\newblock \doi{10.3847/1538-4357/ab4a79}
\bibAnnoteFile{bouffanais2019}

\bibitem[{{Bressan} et~al.(2012){Bressan}, {Marigo}, {Girardi}, {Salasnich},
  {Dal Cero}, {Rubele} et~al.}]{bressan2012}
{Bressan}, A., {Marigo}, P., {Girardi}, L., {Salasnich}, B., {Dal Cero}, C.,
  {Rubele}, S., et~al. (2012).
\newblock {PARSEC: stellar tracks and isochrones with the PAdova and TRieste
  Stellar Evolution Code}.
\newblock \emph{\mnras} 427, 127--145.
\newblock \doi{10.1111/j.1365-2966.2012.21948.x}
\bibAnnoteFile{bressan2012}

\bibitem[{{Burrows} et~al.(2018){Burrows}, {Vartanyan}, {Dolence}, {Skinner},
  and {Radice}}]{burrows2018}
{Burrows}, A., {Vartanyan}, D., {Dolence}, J.~C., {Skinner}, M.~A., and
  {Radice}, D. (2018).
\newblock {Crucial Physical Dependencies of the Core-Collapse Supernova
  Mechanism}.
\newblock \emph{\ssr} 214, 33.
\newblock \doi{10.1007/s11214-017-0450-9}
\bibAnnoteFile{burrows2018}

\bibitem[{{Carr} et~al.(2016){Carr}, {K{\"u}hnel}, and {Sandstad}}]{carr2016}
{Carr}, B., {K{\"u}hnel}, F., and {Sandstad}, M. (2016).
\newblock {Primordial black holes as dark matter}.
\newblock \emph{\prd} 94, 083504.
\newblock \doi{10.1103/PhysRevD.94.083504}
\bibAnnoteFile{carr2016}

\bibitem[{{Chen} et~al.(2014){Chen}, {Woosley}, {Heger}, {Almgren}, and
  {Whalen}}]{chen2014}
{Chen}, K.-J., {Woosley}, S., {Heger}, A., {Almgren}, A., and {Whalen}, D.~J.
  (2014).
\newblock {Two-dimensional Simulations of Pulsational Pair-instability
  Supernovae}.
\newblock \emph{\apj} 792, 28.
\newblock \doi{10.1088/0004-637X/792/1/28}
\bibAnnoteFile{chen2014}

\bibitem[{{Chen} et~al.(2015){Chen}, {Bressan}, {Girardi}, {Marigo}, {Kong},
  and {Lanza}}]{chen2015}
{Chen}, Y., {Bressan}, A., {Girardi}, L., {Marigo}, P., {Kong}, X., and
  {Lanza}, A. (2015).
\newblock {PARSEC evolutionary tracks of massive stars up to 350 M$_{\odot}$ at
  metallicities $0.0001 \le{} Z \le{} 0.04$}.
\newblock \emph{\mnras} 452, 1068--1080.
\newblock \doi{10.1093/mnras/stv1281}
\bibAnnoteFile{chen2015}

\bibitem[{{de Mink} and {Mandel}(2016)}]{demink2016}
{de Mink}, S.~E. and {Mandel}, I. (2016).
\newblock {The chemically homogeneous evolutionary channel for binary black
  hole mergers: rates and properties of gravitational-wave events detectable by
  advanced LIGO}.
\newblock \emph{\mnras} 460, 3545--3553.
\newblock \doi{10.1093/mnras/stw1219}
\bibAnnoteFile{demink2016}

\bibitem[{{Di Carlo} et~al.(2019){Di Carlo}, {Giacobbo}, {Mapelli}, {Pasquato},
  {Spera}, {Wang} et~al.}]{dicarlo2019a}
{Di Carlo}, U.~N., {Giacobbo}, N., {Mapelli}, M., {Pasquato}, M., {Spera}, M.,
  {Wang}, L., et~al. (2019).
\newblock {Merging black holes in young star clusters}.
\newblock \emph{\mnras} 487, 2947--2960.
\newblock \doi{10.1093/mnras/stz1453}
\bibAnnoteFile{dicarlo2019a}

\bibitem[{{Di Carlo} et~al.(2020){Di Carlo}, {Mapelli}, {Bouffanais},
  {Giacobbo}, {Santoliquido}, {Bressan} et~al.}]{dicarlo2019b}
{Di Carlo}, U.~N., {Mapelli}, M., {Bouffanais}, Y., {Giacobbo}, N.,
  {Santoliquido}, F., {Bressan}, A., et~al. (2020).
\newblock {Binary black holes in the pair instability mass gap}.
\newblock \emph{\mnras} 497, 1043--1049.
\newblock \doi{10.1093/mnras/staa1997}
\bibAnnoteFile{dicarlo2019b}

\bibitem[{{Dominik} et~al.(2013){Dominik}, {Belczynski}, {Fryer}, {Holz},
  {Berti}, {Bulik} et~al.}]{dominik2013}
{Dominik}, M., {Belczynski}, K., {Fryer}, C., {Holz}, D.~E., {Berti}, E.,
  {Bulik}, T., et~al. (2013).
\newblock {Double Compact Objects. II. Cosmological Merger Rates}.
\newblock \emph{\apj} 779, 72.
\newblock \doi{10.1088/0004-637X/779/1/72}
\bibAnnoteFile{dominik2013}

\bibitem[{{Eggleton}(2006)}]{eggleton2006}
{Eggleton}, P. (2006).
\newblock \emph{{Evolutionary Processes in Binary and Multiple Stars}}
\bibAnnoteFile{eggleton2006}

\bibitem[{{Ertl} et~al.(2016){Ertl}, {Janka}, {Woosley}, {Sukhbold}, and
  {Ugliano}}]{ertl2016}
{Ertl}, T., {Janka}, H.~T., {Woosley}, S.~E., {Sukhbold}, T., and {Ugliano}, M.
  (2016).
\newblock {A Two-parameter Criterion for Classifying the Explodability of
  Massive Stars by the Neutrino-driven Mechanism}.
\newblock \emph{\apj} 818, 124.
\newblock \doi{10.3847/0004-637X/818/2/124}
\bibAnnoteFile{ertl2016}

\bibitem[{{Farmer} et~al.(2019){Farmer}, {Renzo}, {de Mink}, {Marchant}, and
  {Justham}}]{farmer2019}
{Farmer}, R., {Renzo}, M., {de Mink}, S.~E., {Marchant}, P., and {Justham}, S.
  (2019).
\newblock {Mind the Gap: The Location of the Lower Edge of the Pair-instability
  Supernova Black Hole Mass Gap}.
\newblock \emph{\apj} 887, 53.
\newblock \doi{10.3847/1538-4357/ab518b}
\bibAnnoteFile{farmer2019}

\bibitem[{{Fragos} et~al.(2019){Fragos}, {Andrews}, {Ramirez-Ruiz}, {Meynet},
  {Kalogera}, {Taam} et~al.}]{fragos2019}
{Fragos}, T., {Andrews}, J.~J., {Ramirez-Ruiz}, E., {Meynet}, G., {Kalogera},
  V., {Taam}, R.~E., et~al. (2019).
\newblock {The Complete Evolution of a Neutron-star Binary through a Common
  Envelope Phase Using 1D Hydrodynamic Simulations}.
\newblock \emph{\apjl} 883, L45.
\newblock \doi{10.3847/2041-8213/ab40d1}
\bibAnnoteFile{fragos2019}

\bibitem[{{Fryer}(1999)}]{fryer1999}
{Fryer}, C.~L. (1999).
\newblock {Mass Limits For Black Hole Formation}.
\newblock \emph{\apj} 522, 413--418.
\newblock \doi{10.1086/307647}
\bibAnnoteFile{fryer1999}

\bibitem[{{Fryer} et~al.(2012){Fryer}, {Belczynski}, {Wiktorowicz}, {Dominik},
  {Kalogera}, and {Holz}}]{fryer2012}
{Fryer}, C.~L., {Belczynski}, K., {Wiktorowicz}, G., {Dominik}, M., {Kalogera},
  V., and {Holz}, D.~E. (2012).
\newblock {Compact Remnant Mass Function: Dependence on the Explosion Mechanism
  and Metallicity}.
\newblock \emph{\apj} 749, 91.
\newblock \doi{10.1088/0004-637X/749/1/91}
\bibAnnoteFile{fryer2012}

\bibitem[{{Fryer} and {Kalogera}(2001)}]{fryer2001}
{Fryer}, C.~L. and {Kalogera}, V. (2001).
\newblock {Theoretical Black Hole Mass Distributions}.
\newblock \emph{\apj} 554, 548--560.
\newblock \doi{10.1086/321359}
\bibAnnoteFile{fryer2001}

\bibitem[{{Fuller} and {Ma}(2019)}]{fuller2019}
{Fuller}, J. and {Ma}, L. (2019).
\newblock {Most Black Holes Are Born Very Slowly Rotating}.
\newblock \emph{\apjl} 881, L1.
\newblock \doi{10.3847/2041-8213/ab339b}
\bibAnnoteFile{fuller2019}

\bibitem[{{Gerosa} et~al.(2018){Gerosa}, {Berti}, {O'Shaughnessy},
  {Belczynski}, {Kesden}, {Wysocki} et~al.}]{gerosa2018}
{Gerosa}, D., {Berti}, E., {O'Shaughnessy}, R., {Belczynski}, K., {Kesden}, M.,
  {Wysocki}, D., et~al. (2018).
\newblock {Spin orientations of merging black holes formed from the evolution
  of stellar binaries}.
\newblock \emph{\prd} 98, 084036.
\newblock \doi{10.1103/PhysRevD.98.084036}
\bibAnnoteFile{gerosa2018}

\bibitem[{{Giacobbo} and {Mapelli}(2018)}]{giacobbo2018b}
{Giacobbo}, N. and {Mapelli}, M. (2018).
\newblock {The progenitors of compact-object binaries: impact of metallicity,
  common envelope and natal kicks}.
\newblock \emph{\mnras} 480, 2011--2030.
\newblock \doi{10.1093/mnras/sty1999}
\bibAnnoteFile{giacobbo2018b}

\bibitem[{{Giacobbo} and {Mapelli}(2020)}]{giacobbo2020}
{Giacobbo}, N. and {Mapelli}, M. (2020).
\newblock {Revising Natal Kick Prescriptions in Population Synthesis
  Simulations}.
\newblock \emph{\apj} 891, 141.
\newblock \doi{10.3847/1538-4357/ab7335}
\bibAnnoteFile{giacobbo2020}

\bibitem[{{Giersz} et~al.(2015){Giersz}, {Leigh}, {Hypki}, {L{\"u}tzgendorf},
  and {Askar}}]{giersz2015}
{Giersz}, M., {Leigh}, N., {Hypki}, A., {L{\"u}tzgendorf}, N., and {Askar}, A.
  (2015).
\newblock {MOCCA code for star cluster simulations - IV. A new scenario for
  intermediate mass black hole formation in globular clusters}.
\newblock \emph{\mnras} 454, 3150--3165.
\newblock \doi{10.1093/mnras/stv2162}
\bibAnnoteFile{giersz2015}

\bibitem[{{Gr{\"a}fener} and {Hamann}(2008)}]{graefener2008}
{Gr{\"a}fener}, G. and {Hamann}, W.-R. (2008).
\newblock {Mass loss from late-type WN stars and its Z-dependence. Very massive
  stars approaching the Eddington limit}.
\newblock \emph{\aap} 482, 945--960.
\newblock \doi{10.1051/0004-6361:20066176}
\bibAnnoteFile{graefener2008}

\bibitem[{{Gratton} et~al.(2019){Gratton}, {Bragaglia}, {Carretta}, {D'Orazi},
  {Lucatello}, and {Sollima}}]{gratton2019}
{Gratton}, R., {Bragaglia}, A., {Carretta}, E., {D'Orazi}, V., {Lucatello}, S.,
  and {Sollima}, A. (2019).
\newblock {What is a globular cluster? An observational perspective}.
\newblock \emph{\aapr} 27, 8.
\newblock \doi{10.1007/s00159-019-0119-3}
\bibAnnoteFile{gratton2019}

\bibitem[{{Heger} et~al.(2003){Heger}, {Fryer}, {Woosley}, {Langer}, and
  {Hartmann}}]{heger2003}
{Heger}, A., {Fryer}, C.~L., {Woosley}, S.~E., {Langer}, N., and {Hartmann},
  D.~H. (2003).
\newblock {How Massive Single Stars End Their Life}.
\newblock \emph{\apj} 591, 288--300.
\newblock \doi{10.1086/375341}
\bibAnnoteFile{heger2003}

\bibitem[{{Hills} and {Fullerton}(1980)}]{hills1980}
{Hills}, J.~G. and {Fullerton}, L.~W. (1980).
\newblock {Computer simulations of close encounters between single stars and
  hard binaries}.
\newblock \emph{\aj} 85, 1281--1291.
\newblock \doi{10.1086/112798}
\bibAnnoteFile{hills1980}

\bibitem[{{Hurley} et~al.(2002){Hurley}, {Tout}, and {Pols}}]{hurley2002}
{Hurley}, J.~R., {Tout}, C.~A., and {Pols}, O.~R. (2002).
\newblock {Evolution of binary stars and the effect of tides on binary
  populations}.
\newblock \emph{\mnras} 329, 897--928.
\newblock \doi{10.1046/j.1365-8711.2002.05038.x}
\bibAnnoteFile{hurley2002}

\bibitem[{{Ivanova} et~al.(2013){Ivanova}, {Justham}, {Chen}, {De Marco},
  {Fryer}, {Gaburov} et~al.}]{ivanova2013}
{Ivanova}, N., {Justham}, S., {Chen}, X., {De Marco}, O., {Fryer}, C.~L.,
  {Gaburov}, E., et~al. (2013).
\newblock {Common envelope evolution: where we stand and how we can move
  forward}.
\newblock \emph{\aapr} 21, 59.
\newblock \doi{10.1007/s00159-013-0059-2}
\bibAnnoteFile{ivanova2013}

\bibitem[{{Lada} and {Lada}(2003)}]{lada2003}
{Lada}, C.~J. and {Lada}, E.~A. (2003).
\newblock {Embedded Clusters in Molecular Clouds}.
\newblock \emph{\araa} 41, 57--115.
\newblock \doi{10.1146/annurev.astro.41.011802.094844}
\bibAnnoteFile{lada2003}

\bibitem[{{Mandel} and {de Mink}(2016)}]{mandel2016}
{Mandel}, I. and {de Mink}, S.~E. (2016).
\newblock {Merging binary black holes formed through chemically homogeneous
  evolution in short-period stellar binaries}.
\newblock \emph{\mnras} 458, 2634--2647.
\newblock \doi{10.1093/mnras/stw379}
\bibAnnoteFile{mandel2016}

\bibitem[{{Mapelli}(2016)}]{mapelli2016}
{Mapelli}, M. (2016).
\newblock {Massive black hole binaries from runaway collisions: the impact of
  metallicity}.
\newblock \emph{\mnras} 459, 3432--3446.
\newblock \doi{10.1093/mnras/stw869}
\bibAnnoteFile{mapelli2016}

\bibitem[{{Mapelli}(2018)}]{mapelli2018c}
{Mapelli}, M. (2018).
\newblock {Astrophysics of stellar black holes}.
\newblock \emph{arXiv e-prints} , arXiv:1809.09130
\bibAnnoteFile{mapelli2018c}

\bibitem[{{Mapelli} et~al.(2009){Mapelli}, {Colpi}, and
  {Zampieri}}]{mapelli2009}
{Mapelli}, M., {Colpi}, M., and {Zampieri}, L. (2009).
\newblock {Low metallicity and ultra-luminous X-ray sources in the Cartwheel
  galaxy}.
\newblock \emph{\mnras} 395, L71--L75.
\newblock \doi{10.1111/j.1745-3933.2009.00645.x}
\bibAnnoteFile{mapelli2009}

\bibitem[{{Mapelli} and {Giacobbo}(2018)}]{mapelli2018}
{Mapelli}, M. and {Giacobbo}, N. (2018).
\newblock {The cosmic merger rate of neutron stars and black holes}.
\newblock \emph{\mnras} 479, 4391--4398.
\newblock \doi{10.1093/mnras/sty1613}
\bibAnnoteFile{mapelli2018}

\bibitem[{{Mapelli} et~al.(2017){Mapelli}, {Giacobbo}, {Ripamonti}, and
  {Spera}}]{mapelli2017}
{Mapelli}, M., {Giacobbo}, N., {Ripamonti}, E., and {Spera}, M. (2017).
\newblock {The cosmic merger rate of stellar black hole binaries from the
  Illustris simulation}.
\newblock \emph{\mnras} 472, 2422--2435.
\newblock \doi{10.1093/mnras/stx2123}
\bibAnnoteFile{mapelli2017}

\bibitem[{{Mapelli} et~al.(2010){Mapelli}, {Ripamonti}, {Zampieri}, {Colpi},
  and {Bressan}}]{mapelli2010}
{Mapelli}, M., {Ripamonti}, E., {Zampieri}, L., {Colpi}, M., and {Bressan}, A.
  (2010).
\newblock {Ultra-luminous X-ray sources and remnants of massive metal-poor
  stars}.
\newblock \emph{\mnras} 408, 234--253.
\newblock \doi{10.1111/j.1365-2966.2010.17048.x}
\bibAnnoteFile{mapelli2010}

\bibitem[{{Mapelli} et~al.(2020){Mapelli}, {Spera}, {Montanari}, {Limongi},
  {Chieffi}, {Giacobbo} et~al.}]{mapelli2020}
{Mapelli}, M., {Spera}, M., {Montanari}, E., {Limongi}, M., {Chieffi}, A.,
  {Giacobbo}, N., et~al. (2020).
\newblock {Impact of the Rotation and Compactness of Progenitors on the Mass of
  Black Holes}.
\newblock \emph{\apj} 888, 76.
\newblock \doi{10.3847/1538-4357/ab584d}
\bibAnnoteFile{mapelli2020}

\bibitem[{{Marchant} et~al.(2016){Marchant}, {Langer}, {Podsiadlowski},
  {Tauris}, and {Moriya}}]{marchant2016}
{Marchant}, P., {Langer}, N., {Podsiadlowski}, P., {Tauris}, T.~M., and
  {Moriya}, T.~J. (2016).
\newblock {A new route towards merging massive black holes}.
\newblock \emph{\aap} 588, A50.
\newblock \doi{10.1051/0004-6361/201628133}
\bibAnnoteFile{marchant2016}

\bibitem[{{Marchant} et~al.(2019){Marchant}, {Renzo}, {Farmer}, {Pappas},
  {Taam}, {de Mink} et~al.}]{marchant2019}
{Marchant}, P., {Renzo}, M., {Farmer}, R., {Pappas}, K. M.~W., {Taam}, R.~E.,
  {de Mink}, S.~E., et~al. (2019).
\newblock {Pulsational Pair-instability Supernovae in Very Close Binaries}.
\newblock \emph{\apj} 882, 36.
\newblock \doi{10.3847/1538-4357/ab3426}
\bibAnnoteFile{marchant2019}

\bibitem[{{Neijssel} et~al.(2019){Neijssel}, {Vigna-G{\'o}mez}, {Stevenson},
  {Barrett}, {Gaebel}, {Broekgaarden} et~al.}]{neijssel2019}
{Neijssel}, C.~J., {Vigna-G{\'o}mez}, A., {Stevenson}, S., {Barrett}, J.~W.,
  {Gaebel}, S.~M., {Broekgaarden}, F.~S., et~al. (2019).
\newblock {The effect of the metallicity-specific star formation history on
  double compact object mergers}.
\newblock \emph{\mnras} 490, 3740--3759.
\newblock \doi{10.1093/mnras/stz2840}
\bibAnnoteFile{neijssel2019}

\bibitem[{{Neumayer} et~al.(2020){Neumayer}, {Seth}, and
  {B{\"o}ker}}]{neumayer2020}
{Neumayer}, N., {Seth}, A., and {B{\"o}ker}, T. (2020).
\newblock {Nuclear star clusters}.
\newblock \emph{\aapr} 28, 4.
\newblock \doi{10.1007/s00159-020-00125-0}
\bibAnnoteFile{neumayer2020}

\bibitem[{{Ober} et~al.(1983){Ober}, {El Eid}, and {Fricke}}]{ober1983}
{Ober}, W.~W., {El Eid}, M.~F., and {Fricke}, K.~J. (1983).
\newblock {Evolution of Massive Pregalactic Stars - Part Two - Nucleosynthesis
  in Pair Creation Supernovae and Pregalactic Enrichment}.
\newblock \emph{\aap} 119, 61
\bibAnnoteFile{ober1983}

\bibitem[{{O'Connor} and {Ott}(2011)}]{oconnor2011}
{O'Connor}, E. and {Ott}, C.~D. (2011).
\newblock {Black Hole Formation in Failing Core-Collapse Supernovae}.
\newblock \emph{\apj} 730, 70.
\newblock \doi{10.1088/0004-637X/730/2/70}
\bibAnnoteFile{oconnor2011}

\bibitem[{{Portegies Zwart} et~al.(2004){Portegies Zwart}, {Baumgardt}, {Hut},
  {Makino}, and {McMillan}}]{portegieszwart2004}
{Portegies Zwart}, S.~F., {Baumgardt}, H., {Hut}, P., {Makino}, J., and
  {McMillan}, S.~L.~W. (2004).
\newblock {Formation of massive black holes through runaway collisions in dense
  young star clusters}.
\newblock \emph{\nat} 428, 724--726.
\newblock \doi{10.1038/nature02448}
\bibAnnoteFile{portegieszwart2004}

\bibitem[{{Portegies Zwart} and {McMillan}(2000)}]{portegieszwart2000}
{Portegies Zwart}, S.~F. and {McMillan}, S.~L.~W. (2000).
\newblock {Black Hole Mergers in the Universe}.
\newblock \emph{\apjl} 528, L17--L20.
\newblock \doi{10.1086/312422}
\bibAnnoteFile{portegieszwart2000}

\bibitem[{{Portegies Zwart} et~al.(2010){Portegies Zwart}, {McMillan}, and
  {Gieles}}]{portegieszwart2010}
{Portegies Zwart}, S.~F., {McMillan}, S.~L.~W., and {Gieles}, M. (2010).
\newblock {Young Massive Star Clusters}.
\newblock \emph{\araa} 48, 431--493.
\newblock \doi{10.1146/annurev-astro-081309-130834}
\bibAnnoteFile{portegieszwart2010}

\bibitem[{{Qin} et~al.(2018){Qin}, {Fragos}, {Meynet}, {Andrews},
  {S{\o}rensen}, and {Song}}]{qin2018}
{Qin}, Y., {Fragos}, T., {Meynet}, G., {Andrews}, J., {S{\o}rensen}, M., and
  {Song}, H.~F. (2018).
\newblock {The spin of the second-born black hole in coalescing binary black
  holes}.
\newblock \emph{\aap} 616, A28.
\newblock \doi{10.1051/0004-6361/201832839}
\bibAnnoteFile{qin2018}

\bibitem[{{Qin} et~al.(2019){Qin}, {Marchant}, {Fragos}, {Meynet}, and
  {Kalogera}}]{qin2019}
{Qin}, Y., {Marchant}, P., {Fragos}, T., {Meynet}, G., and {Kalogera}, V.
  (2019).
\newblock {On the Origin of Black Hole Spin in High-mass X-Ray Binaries}.
\newblock \emph{\apjl} 870, L18.
\newblock \doi{10.3847/2041-8213/aaf97b}
\bibAnnoteFile{qin2019}

\bibitem[{{Renzo} et~al.(2020){Renzo}, {Farmer}, {Justham}, {de Mink},
  {G{\"o}tberg}, and {Marchant}}]{renzo2020}
{Renzo}, M., {Farmer}, R.~J., {Justham}, S., {de Mink}, S.~E., {G{\"o}tberg},
  Y., and {Marchant}, P. (2020).
\newblock {Sensitivity of the lower edge of the pair-instability black hole
  mass gap to the treatment of time-dependent convection}.
\newblock \emph{\mnras} 493, 4333--4341.
\newblock \doi{10.1093/mnras/staa549}
\bibAnnoteFile{renzo2020}

\bibitem[{{Rodriguez} et~al.(2018){Rodriguez}, {Amaro-Seoane}, {Chatterjee},
  and {Rasio}}]{rodriguez2018}
{Rodriguez}, C.~L., {Amaro-Seoane}, P., {Chatterjee}, S., and {Rasio}, F.~A.
  (2018).
\newblock {Post-Newtonian Dynamics in Dense Star Clusters: Highly Eccentric,
  Highly Spinning, and Repeated Binary Black Hole Mergers}.
\newblock \emph{\prl} 120, 151101.
\newblock \doi{10.1103/PhysRevLett.120.151101}
\bibAnnoteFile{rodriguez2018}

\bibitem[{{Rodriguez} et~al.(2019){Rodriguez}, {Zevin}, {Amaro-Seoane},
  {Chatterjee}, {Kremer}, {Rasio} et~al.}]{rodriguez2019}
{Rodriguez}, C.~L., {Zevin}, M., {Amaro-Seoane}, P., {Chatterjee}, S.,
  {Kremer}, K., {Rasio}, F.~A., et~al. (2019).
\newblock {Black holes: The next generation{\textemdash}repeated mergers in
  dense star clusters and their gravitational-wave properties}.
\newblock \emph{\prd} 100, 043027.
\newblock \doi{10.1103/PhysRevD.100.043027}
\bibAnnoteFile{rodriguez2019}

\bibitem[{{Rodriguez} et~al.(2016){Rodriguez}, {Zevin}, {Pankow}, {Kalogera},
  and {Rasio}}]{rodriguez2016b}
{Rodriguez}, C.~L., {Zevin}, M., {Pankow}, C., {Kalogera}, V., and {Rasio},
  F.~A. (2016).
\newblock {Illuminating Black Hole Binary Formation Channels with Spins in
  Advanced LIGO}.
\newblock \emph{\apjl} 832, L2.
\newblock \doi{10.3847/2041-8205/832/1/L2}
\bibAnnoteFile{rodriguez2016b}

\bibitem[{{Samsing}(2018)}]{samsing2018}
{Samsing}, J. (2018).
\newblock {Eccentric black hole mergers forming in globular clusters}.
\newblock \emph{\prd} 97, 103014.
\newblock \doi{10.1103/PhysRevD.97.103014}
\bibAnnoteFile{samsing2018}

\bibitem[{{Samsing} et~al.(2014){Samsing}, {MacLeod}, and
  {Ramirez-Ruiz}}]{samsing2014}
{Samsing}, J., {MacLeod}, M., and {Ramirez-Ruiz}, E. (2014).
\newblock {The Formation of Eccentric Compact Binary Inspirals and the Role of
  Gravitational Wave Emission in Binary-Single Stellar Encounters}.
\newblock \emph{\apj} 784, 71.
\newblock \doi{10.1088/0004-637X/784/1/71}
\bibAnnoteFile{samsing2014}

\bibitem[{{Santoliquido} et~al.(2020){Santoliquido}, {Mapelli}, {Bouffanais},
  {Giacobbo}, {Di Carlo}, {Rastello} et~al.}]{santoliquido2020}
{Santoliquido}, F., {Mapelli}, M., {Bouffanais}, Y., {Giacobbo}, N., {Di
  Carlo}, U.~N., {Rastello}, S., et~al. (2020).
\newblock {The Cosmic Merger Rate Density Evolution of Compact Binaries Formed
  in Young Star Clusters and in Isolated Binaries}.
\newblock \emph{\apj} 898, 152.
\newblock \doi{10.3847/1538-4357/ab9b78}
\bibAnnoteFile{santoliquido2020}

\bibitem[{{Spera} and {Mapelli}(2017)}]{spera2017}
{Spera}, M. and {Mapelli}, M. (2017).
\newblock {Very massive stars, pair-instability supernovae and
  intermediate-mass black holes with the sevn code}.
\newblock \emph{\mnras} 470, 4739--4749.
\newblock \doi{10.1093/mnras/stx1576}
\bibAnnoteFile{spera2017}

\bibitem[{{Spera} et~al.(2019){Spera}, {Mapelli}, {Giacobbo}, {Trani},
  {Bressan}, and {Costa}}]{spera2019}
{Spera}, M., {Mapelli}, M., {Giacobbo}, N., {Trani}, A.~A., {Bressan}, A., and
  {Costa}, G. (2019).
\newblock {Merging black hole binaries with the SEVN code}.
\newblock \emph{\mnras} 485, 889--907.
\newblock \doi{10.1093/mnras/stz359}
\bibAnnoteFile{spera2019}

\bibitem[{{Spitzer}(1987)}]{spitzer1987}
{Spitzer}, L. (1987).
\newblock \emph{{Dynamical evolution of globular clusters}}
\bibAnnoteFile{spitzer1987}

\bibitem[{{Stevenson} et~al.(2019){Stevenson}, {Sampson}, {Powell},
  {Vigna-G{\'o}mez}, {Neijssel}, {Sz{\'e}csi} et~al.}]{stevenson2019}
{Stevenson}, S., {Sampson}, M., {Powell}, J., {Vigna-G{\'o}mez}, A.,
  {Neijssel}, C.~J., {Sz{\'e}csi}, D., et~al. (2019).
\newblock {The Impact of Pair-instability Mass Loss on the Binary Black Hole
  Mass Distribution}.
\newblock \emph{\apj} 882, 121.
\newblock \doi{10.3847/1538-4357/ab3981}
\bibAnnoteFile{stevenson2019}

\bibitem[{{Tang} et~al.(2020){Tang}, {Eldridge}, {Stanway}, and
  {Bray}}]{tang2020}
{Tang}, P.~N., {Eldridge}, J.~J., {Stanway}, E.~R., and {Bray}, J.~C. (2020).
\newblock {Dependence of gravitational wave transient rates on cosmic star
  formation and metallicity evolution history}.
\newblock \emph{\mnras} 493, L6--L10.
\newblock \doi{10.1093/mnrasl/slz183}
\bibAnnoteFile{tang2020}

\bibitem[{{Vink} et~al.(2001){Vink}, {de Koter}, and {Lamers}}]{vink2001}
{Vink}, J.~S., {de Koter}, A., and {Lamers}, H.~J.~G.~L.~M. (2001).
\newblock {Mass-loss predictions for O and B stars as a function of
  metallicity}.
\newblock \emph{\aap} 369, 574--588.
\newblock \doi{10.1051/0004-6361:20010127}
\bibAnnoteFile{vink2001}

\bibitem[{{Vink} et~al.(2011){Vink}, {Muijres}, {Anthonisse}, {de Koter},
  {Gr{\"a}fener}, and {Langer}}]{vink2011}
{Vink}, J.~S., {Muijres}, L.~E., {Anthonisse}, B., {de Koter}, A.,
  {Gr{\"a}fener}, G., and {Langer}, N. (2011).
\newblock {Wind modelling of very massive stars up to 300 solar masses}.
\newblock \emph{\aap} 531, A132.
\newblock \doi{10.1051/0004-6361/201116614}
\bibAnnoteFile{vink2011}

\bibitem[{{Woosley}(2017)}]{woosley2017}
{Woosley}, S.~E. (2017).
\newblock {Pulsational Pair-instability Supernovae}.
\newblock \emph{\apj} 836, 244.
\newblock \doi{10.3847/1538-4357/836/2/244}
\bibAnnoteFile{woosley2017}

\bibitem[{{Woosley}(2019)}]{woosley2019}
{Woosley}, S.~E. (2019).
\newblock {The Evolution of Massive Helium Stars, Including Mass Loss}.
\newblock \emph{\apj} 878, 49.
\newblock \doi{10.3847/1538-4357/ab1b41}
\bibAnnoteFile{woosley2019}

\bibitem[{{Woosley} et~al.(2007){Woosley}, {Blinnikov}, and
  {Heger}}]{woosley2007}
{Woosley}, S.~E., {Blinnikov}, S., and {Heger}, A. (2007).
\newblock {Pulsational pair instability as an explanation for the most luminous
  supernovae}.
\newblock \emph{\nat} 450, 390--392.
\newblock \doi{10.1038/nature06333}
\bibAnnoteFile{woosley2007}

\bibitem[{{Yoshida} et~al.(2016){Yoshida}, {Umeda}, {Maeda}, and
  {Ishii}}]{yoshida2016}
{Yoshida}, T., {Umeda}, H., {Maeda}, K., and {Ishii}, T. (2016).
\newblock {Mass ejection by pulsational pair instability in very massive stars
  and implications for luminous supernovae}.
\newblock \emph{\mnras} 457, 351--361.
\newblock \doi{10.1093/mnras/stv3002}
\bibAnnoteFile{yoshida2016}

\bibitem[{{Zampieri} and {Roberts}(2009)}]{zampieri2009}
{Zampieri}, L. and {Roberts}, T.~P. (2009).
\newblock {Low-metallicity natal environments and black hole masses in
  ultraluminous X-ray sources}.
\newblock \emph{\mnras} 400, 677--686.
\newblock \doi{10.1111/j.1365-2966.2009.15509.x}
\bibAnnoteFile{zampieri2009}

\bibitem[{{Zevin} et~al.(2017){Zevin}, {Pankow}, {Rodriguez}, {Sampson},
  {Chase}, {Kalogera} et~al.}]{zevin2017}
{Zevin}, M., {Pankow}, C., {Rodriguez}, C.~L., {Sampson}, L., {Chase}, E.,
  {Kalogera}, V., et~al. (2017).
\newblock {Constraining Formation Models of Binary Black Holes with
  Gravitational-wave Observations}.
\newblock \emph{\apj} 846, 82.
\newblock \doi{10.3847/1538-4357/aa8408}
\bibAnnoteFile{zevin2017}

\bibitem[{{Zevin} et~al.(2019){Zevin}, {Samsing}, {Rodriguez}, {Haster}, and
  {Ramirez-Ruiz}}]{zevin2019}
{Zevin}, M., {Samsing}, J., {Rodriguez}, C., {Haster}, C.-J., and
  {Ramirez-Ruiz}, E. (2019).
\newblock {Eccentric Black Hole Mergers in Dense Star Clusters: The Role of
  Binary-Binary Encounters}.
\newblock \emph{\apj} 871, 91.
\newblock \doi{10.3847/1538-4357/aaf6ec}
\bibAnnoteFile{zevin2019}

\bibitem[{{Ziosi} et~al.(2014){Ziosi}, {Mapelli}, {Branchesi}, and
  {Tormen}}]{ziosi2014}
{Ziosi}, B.~M., {Mapelli}, M., {Branchesi}, M., and {Tormen}, G. (2014).
\newblock {Dynamics of stellar black holes in young star clusters with
  different metallicities - II. Black hole-black hole binaries}.
\newblock \emph{\mnras} 441, 3703--3717.
\newblock \doi{10.1093/mnras/stu824}
\bibAnnoteFile{ziosi2014}

\end{thebibliography}


\end{document}